# Superconductivity with Topological non-trivial surface states in NbC

N.K. Karn[1,2], M.M. Sharma[1,2], Prince Sharma[1,2], Ganesh Gurjar[3], S. Patnaik[3], and V.P.S. Awana[1,2, *]

*[1]Academy of Scientific & Innovative Research (AcSIR), Ghaziabad-201002*

*[2]CSIR- National Physical Laboratory, New Delhi-110012*

*[3] School of Physical Sciences, Jawaharlal Nehru University, New Delhi-110067*

**Abstract:**

Quantum materials with non-trivial band topology and bulk superconductivity are considered superior materials to realize topological superconductivity. In this regard, we report detailed Density Functional Theory (DFT) calculations and Z2 invaraints for the NbC superconductor, exhibiting its band structure to be topologically non-trivial. Bulk superconductivity at 8.9K is confirmed through DC magnetization measurements under Field Cooled (FC) and Zero Field Cooled (ZFC) protocols. This superconductivity is found to be of type-II nature as revealed by isothermal M-H measurements and thus calculated the Ginzberg-Landau parameter. A large intermediate state is evident from the phase diagram, showing NbC to be a strong type-II superconductor. Comparing with earlier reports on superconducting NbC, a non-monotonic relationship of critical temperature with lattice parameters is seen. In conclusion, NbC is a type-II around 10K superconductor with topological non-trivial surface states.

*Corresponding Author

Dr. V. P. S. Awana:  E-mail: awana@nplindia.org
Ph. +91-11-45609357, Fax-+91-11-45609310
Homepage: awanavps.webs.com

**Introduction:**

Condensed matter scientists are always keen to discover new materials and explore their novel physical properties [1,2]. In this regard, materials with topologically non-trivial band structures have proved to be a rich exploring field for condensed matter physicists [1,2]. These materials show a variety of properties, such as high magneto-resistance (MR) [3,4], high surface conductivity [5], and Terahertz (THz) generation [6,7]. Coming to the superconducting properties, these topological materials with non-trivial band structure form a new class of materials named; Topological superconductors (TSc) [2]. TSc has a superconducting gap in the bulk and topologically non-trivial states at the surface [8-10]. This unique feature of TSc makes them the most promising candidates to host the Majorana Fermions [9]. Majorana Fermions are the anti-particle of themselves, which follow the Dirac relativistic equation. The existence of Majorana Fermions qualifies the TSc to be the essential quantum material for fault-tolerant quantum computing [11]. It is thus essential to search for materials that show topological superconductivity in bulk form.

Topological Superconductivity in bulk material can be induced either by doping suitable elements in topological material [12–16] or by applying pressure [17,18]. Apart from these methods, some Dirac type-II semimetals show superconductivity in their intrinsic form [19,20]. Dirac semimetals (DSM) have a Dirac point at which bands are four-fold degenerate. DSM is categorized as type-I DSM and type-II DSM based on the nature of their Dirac point. The Dirac point in type-I DSM is point-like, while in the case of Type-II DSM, it is tilted at some angle [21]. Recently, some Transition metal carbides (TMCs) viz. TaC [22,23], NbC [23,24] are found to have DSM-like band structures along with bulk superconductivity. Superconductivity of these materials was observed a long time ago and is summarized in a combined report on the Superconductivity of TMCs [25], but their topologically non-trivial character was not explored. Other TMCs, such as VC [26] and CrC [26], are also predicted to have non-trivial band topology along with bulk superconductivity, but experimental confirmation of their non-trivial topological behavior is yet to be realized. Among these TMCs, the NbC and TaC show superconductivity in the range of 8 to 11K [22–24,27–29]. Most of the Superconductivity reports of these materials are based on polycrystalline samples, as it is challenging to grow their single crystals. The superconducting critical temperature ($T_c$) of polycrystalline NbC varies with the lattice parameter [28] due to variation in the stoichiometry of Carbon [30,31]. Non-trivial band topology in NbC was recently predicated theoretically through DFT calculations [23,24]. In a recent report on single crystalline NbC being synthesized using Co as flux [24], the Angle-Resolved Photoelectron Spectroscopy (ARPES) measurements showed clear Fermi surface nesting origin of non-trivial band topology in NbC.

In this work, we synthesize a polycrystalline sample of NbC in a single step at a relatively higher temperature ($1350^0$C) from our previous report on NbC [28]. The polycrystalline sample has a single-phase and shows type-II superconductivity at below 9K. Band structure calculations are also performed by using DFT at the same path as suggested in ref.24. It is found that NbC possesses the non-trivial band topology along with 8.9 K superconductivity. However, it has been proved in ref. 28 that $T_c$ and lattice parameters strongly depend on synthesis temperature and increase as the synthesis temperature increases. But, in this report, it does not follow the same trend. Surprisingly, lattice parameters are increased from optimum value as the synthesis temperature increases from that previously reported. However, the $T_c$ is decreased and showing a non-monotonic relationship with lattice parameters and synthesis temperature. This particular ambiguity is highlighted in this article, along with an in-depth analysis of band structure and projected DOS (density of states) in the 12 X 12 X 12 matrix including calculation of Z2 invariants. Our results, along with [23, 24], unambiguously prove that NbC is a 10K superconductor [28, 30, 31] with non-trivial band topology.

**Experimental:**

A polycrystalline sample of NbC is prepared in a single step by following the solid-state reaction route as suggested in ref. [28], but at a slightly elevated temperature. High purity (>4N) powders of Nb and amorphous Carbon were taken into stoichiometric ratio. These powders were mixed and ground thoroughly by using agate mortar pestle in an argon-filled Glove Box. This homogenous mixture was then palletized and vacuum encapsulated at a pressure of $5\times10^{-5}$ mbar. This vacuum encapsulated sample was then placed into an automated PID controlled Muffle furnace and heated to $1350^0$C at a rate of $120^0$C/h. This sample is kept at this elevated temperature for a period of 48h. After this, the sample is allowed to cool normally to room temperature.

The PXRD pattern of synthesized polycrystalline NbC crystal was recorded using Rigaku mini flex-II tabletop X-ray diffractometer equipped with CuK$_\alpha$ radiation of 1.5418Å wavelength. Rietveld refinement of the PXRD pattern was performed using Full Proof software, and the Unit cell of NbC was drawn using VESTA software. Band structure calculations were performed under DFT protocols using Quantum Espresso (QE) software in QUANTUM VITAS [33,34], whose optimized path is considered from the ARPES study in ref. 24. Moreover, in order to calculate the Z2 invariants of the NbC, *WANNIER90* is used in which wannierization of Bloch wave function is implemented [35]. In this regard, the MLWFs (Maximally localized Wannnier Function) are considered, which is used to verify the band structure that is generated through the first principle method. Based on the MLWFs, we obtain an effective tight-binding (TB) model for the system NbC. This TB model is further processed and implemented in WANNIER-TOOLS [36] on a 4×4×4 K-mesh, which samples the whole Brillouin Zone. Wanni*er* charge centers are further calculated of which evolution in Brillouin zone planes indicated the states of Z2-invariant. Magnetization measurements were carried out on Quantum Design Physical Property Measurement System (PPMS).

**Results & Discussion:**

Fig. 1 shows Rietveld's refined PXRD pattern of synthesized polycrystalline powder of NbC, confirming the single phase of the material, as no impurity peak is observed in the PXRD. Synthesized NbC sample has a cubic structure with F m -3 m space group symmetry. Rietveld refined PXRD pattern shows that all observed peaks are well fitted under F m -3 m space group symmetry. The goodness of fit $\chi^2$ is found to be 2.85, which is in an acceptable range. The constituent elements; Nb and C occupy (0, 0, 0) and (0, 0, 0.5) atomic positions in the lattice. Lattice parameters obtained from Rietveld refinement are a=b=c= 4.4739(8) Å & $\alpha=\beta=\gamma= 90^0$. The obtained lattice parameters are slightly higher than as reported in our previous report [28]. It was observed earlier [28] that lattice parameters depend on the temperature at which the sample is prepared, and the same increases with the processing temperature. Here, the obtained results are consistent with this fact as the sample is prepared at a higher temperature than reported in ref. 28. The variation in Lattice parameters in NbC results is due to the change in the stoichiometry of Carbon [30,31]. It has been suggested in previous reports that the lattice parameter of NbC increases as the Carbon content increases. For exact stoichiometric NbC, lattice parameters are observed to be 4.4704 Å, and all previous reports are based on Carbon deficient NbC, for which the lattice parameter lies between 4.4281 Å to 4.4704 Å (for NbC) [30]. Here, the lattice parameter is greater than that is observed for exact stoichiometric NbC. It suggests that the synthesized NbC sample has a C/Nb ratio greater than 1. The XRD pattern cannot determine the exact carbon content in the synthesized NbC sample as Carbon is a lighter element. All empirical formulas that are reported to calculate stoichiometry of carbon in NbC work only from 0.7 to 0.96 molar ratio of C with respect to Nb. It is challenging to predict how much Carbon content is there in the sample, but the enhancement of lattice parameters from NbC suggests that the Carbon has a higher molar ratio with respect to Nb in presently synthesized NbC polycrystalline sample.

Fig. 2 shows FC and ZFC measurements of presently synthesized NbC polycrystalline samples in a magnetic field of 8Oe. These results show a clear diamagnetic transition in both FC and ZFC measurements at 8.9K. This diamagnetic transition signifies the presence of bulk superconductivity in synthesized NbC. The superconducting volume fraction is around 31% in ZFC measurements which is quite good for polycrystalline samples. It signifies that the sample is crystallized in a single phase. This ZFC signal starts to saturate near 6K, showing transition width to be around 3K. The observed superconducting transition is lesser than the optimum value, i.e., 11.5K. The reason behind this lower $T_c$ is the stoichiometry of Carbon which eventually changes the lattice parameter. From the previous reports, it is well established that $T_c$ in NbC strongly depends on Carbon stoichiometry and lattice parameters [28,30]. Our previous report also showed the same where it is concluded that the lattice parameters change with the synthesis temperature, and accordingly, the $T_c$ also changes [28]. Maximum $T_c$ was achieved in the sample prepared at $1250^0$C. However, the sample is synthesized here is at an even higher temperature of $1350^0$C. The growth temperature increases the lattice parameter from the perfectly stoichiometric NbC. It signifies that Carbon content is higher than that for Nb. So, it can be said that both Carbon deficient and Carbon-rich NbC phases have lower $T_c$ than the optimum value, which is indirectly influenced by the lattice parameters. In the inset of fig. 2(a), the dependence of superconducting transitions of NbC samples on their lattice parameters is highlighted. The critical temperature for NbC samples synthesized at a lower temperature is taken from ref. [28]. It shows that the optimum $T_c$ is obtained at $1250^0$C with the exact stoichiometric composition of NbC, and if the processing temperature is further increased, it results in a decrement in $T_c$ due to increment in Carbon content as seen for present NbC, being synthesized at $1350^0$C. It shows that $T_c$ has a non-monotonic relationship with the lattice parameter in NbC. Lattice parameters and $T_c$ of NbC samples synthesized at $1350^o$C and at lower temperatures in ref. [28], are summarized in Table 1.

Fig. 2(b) shows isothermal M-H plots of synthesized NbC samples at 2K, 4K, 6K, 8K, and 10K. These plots start with linear variation in magnetization with respect to the applied field, showing Meissner's state. Then after a critical field, it deviates from this linearity and enters into a mixed state. It is a clear signature of type-II superconductivity in synthesized NbC samples. The field at which the M-H plot starts to deviate from linearity is known as the lower critical field $H_{c1}$. While the upper critical field $H_{c2}$ is the field at which the M-H plot coincides with the irreversibility of the field. The M-H plots are wide opened up to 8K while it is linear at 10K. This result is consistent with FC and ZFC measurements which show that superconductivity persists up to 8.9K. Inset of fig. 2(b) shows the M-H plot at 2K. Upper critical field $H_{c2}$ is marked at which the M-H plot becomes irreversible, and it is found to be at 23kOe. From the M-H plot in fig. 2(b) the value of upper critical field $H_{c2}$ at 4K, 6K, and 8K are found to be 19kOe, 12kOe, and 3.6kOe, respectively.

Lower critical field $H_{c1}$ is the field at which the M-H plot deviates from linearity, or this is the field at which the first magnetic line of force penetrates the superconducting sample. Fig. 2(c) shows M-H plots of synthesized NbC at 2K, 4K, and 6K in the low magnetic field region. To get an estimate of $H_{c1}$, a straight line is marked, which is showing the deviation from linearity. However, from this, the accurate value of $H_{c1}$ cannot be determined. A separate exercise described in ref. [32] is made to determine the value of $H_{c1}$. In this method, the intention is to determine the point at which the M-H loop deviates from linearity and enters a mixed state. For this, first, the low magnetic field M-H data is linearly fitted. The slope of this linear fit is used to determine $M_0$, and then this $M_0$ is subtracted from each isotherm. This difference (M-$M_0$) is represented as $\Delta M$, and this change is plotted against the applied field. A baseline is drawn for $\Delta M = 0$, and $H_{c1}$ is the point at which $\Delta M$ deviates from this zero baseline. This exercise is shown in the inset of Fig. 2(c). The values of $H_{c1}$ are found to be 81 Oe, 69 Oe, and 60 Oe at 2K, 4K, and 6K, respectively. There is a big difference between the values of $H_{c1}$ and $H_{c2}$ at 2K. It signifies that mixed states persist for approx. 22kOe and synthesized NbC sample is a strong type-II superconductor.

The values of $H_{c1}$ and $H_{c2}$ at different temperatures viz. 2K, 4K, 6K, and 8K are used to plot the phase diagram of the synthesized NbC polycrystalline sample, as shown in Fig. 2(d). This phase diagram is clearly showing the superconducting state, intermediate state, and the normal state of NbC. This phase diagram is plotted by fitting the values of $H_{c1}$ and $H_{c2}$ with the following equations

$$H_{c1}(T) = H_{c1}(0) \ [1 - T^2/T_c^2] \ \& \ H_{c2}(T) = H_{c2}(0) \ [1 - T^2/T_c^2]$$

The values of $H_{c1}$ and $H_{c2}$ are well fitted with their model equations which are standard behavior for superconductors. The phase diagram is plotted by taking the Y-axis as the difference between the values of $H_{c1}$ and $H_{c2}$ is quite large. The left and right-hand side axis are showing the values of $H_{c1}$ and $H_{c2}$, respectively. A relatively large intermediate state can be clearly seen in the phase diagram of NbC, which confirms it to be a strong type-II superconductor.

The values of $H_{c1}$ and $H_{c2}$ are used to determine the mean critical field ($H_c$) at 2K by using the $H_c = (H_{c1}*H_{c2})^{1/2}$. The values of $H_{c1}$ and $H_{c2}$ at 2K are 81 Oe and 23 kOe, respectively. Using these values, $H_c$ is found to be 1.36 kOe. It means the critical field will be used to determine the value of the upper critical field at absolute zero $H_{c2}(0)$. $H_{c2}(0)$ is calculated by using the Ginzberg Landau equation (G-L equation), which is given by

$$H_{c2}(T) = H_{c2}(0) * \left[ \frac{1 - t^2}{1 + t^2} \right]$$

Here, in the above equation, $t = T/T_c$ known as reduced temperature. T is the temperature at which the upper critical field is taken, i.e., 2K and $T_c$ is critical temperature, taken to be 8.9K as observed in FC and ZFC measurements, giving a reduced temperature of 0.225. So, by using these values, $H_{c2}(0)$ is found to be 25.42kOe. The nature of superconductivity, whether type-I or type-II, can be determined by calculating the G-L kappa ($\kappa$) parameter. The $\kappa$ parameter can be calculated by following the relation $H_{c2}(0) = \kappa*(2)^{1/2}*H_c$. The value of $\kappa$ parameter is found to be 13.21. This value is much above the threshold value for type-I superconductivity, suggesting that the synthesized NbC sample is a strong type-II superconductor, supporting the wide opened M-H plots in Fig 2(b) too. Other critical parameters of superconductivity, such as coherence length $\xi(0)$, can be calculated using the formula

$$H_{c2}(0) = \frac{\varphi_0}{2 \Pi \xi(0)^2}$$

Where $\Phi_0$, a constant term is known as flux quanta. The value of this constant term is 2.0678 x $10^{-15}$ Wb. From the above equation, the value of $\xi(0)$ is found to be 1.29Å. Penetration depth $\lambda(0)$ can be calculated by using the relation $\kappa = \lambda(0)/\xi(0)$, and it is found to be 17.04Å.

Further, to determine the projected DOS and possible topological non-triviality of the band structure of synthesized NbC polycrystalline sample, the DFT calculations are performed by using software QUANTUM VITAS which depends on QUANTUM EXPRESSO. The CIF (crystallographic information framework/file) produced through Rietveld refinement is used to calculate the band structure. DFT calculations are performed by considering both with and without Spin-Orbit Coupling using the crystal parameters from the grown NbC. The k-path which is taken to calculate band structure is as follows:

$$W \rightarrow L \rightarrow K \rightarrow \Gamma \rightarrow X$$

This particular K-path is considered from ref.24. The calculation are realized in Quantum Espresso with Perdew-Burke-Ernzerhof (PBE) exchange-correlation functional [33,34]. The calculated band structure along with projected DOS with and without SOC are shown in fig. 3(a) and 3(b), respectively. Fermi level is marked as zero energy. It is clear from the projected DOS plots that bands near the Fermi level are generated due to the hybridization of 2p orbitals of Carbon and 4d orbitals of Nb. The d orbitals of Nb dominate in these bands. Finite DOS present near the Fermi level reveals the metallic nature of synthesized NbC polycrystalline sample too. The splitting of p and d orbitals bands is observed with the inclusion of SOC, as shown in fig. 3(b). This particular splitting confirms the presence of effective SOC in the NbC. The calculated band structure is shown in the right-hand side image of Fig. 3(a) and 3(b).

Most of the bands are shown to cross the Fermi level, which confirms the metallic nature of the synthesized polycrystalline NbC sample. A Dirac cone-like structure can be visualized at $\Gamma$ point in the Brillouin zone. The enlarged view of bands near this point is shown in Fig. 3(c). A similar Dirac cone-like band structure is observed near this $\Gamma$, which is encircled in Fig. 3(a). This Dirac cone-like structure is found to be tilted at some angle, and bands observed at $\Gamma$ point are six-fold degenerate. Band structure with inclusion of SOC is shown in Fig. 3(b). It has been observed that after the inclusion of SOC, the six-fold degenerate bands observed at $\Gamma$ point in without SOC plots become fourfold degenerate. It can be seen in Fig. 3(c), which contains an enlarged view of band structure too around Γ point with and without SOC. It suggests that the degeneracy of bands is lifted when SOC is included. The Dirac cone-like structure encircled in Fig. 3(a) also split into several bands when SOC is included. It can be seen in the inset of Fig. 3(b), which shows that all bands become gapped when SOC is included. These results are in good agreement with the previous theoretical calculation made on this compound [23,24]. It suggests that SOC is effective in NbC samples. All the bands are found to be splitted at and around Γ point, when SOC is included, and are observed near the Fermi level, suggesting that the synthesized NbC sample has a topologically non-trivial band structure. The tilted Dirac cone-like structure observed near Γ point suggests that the NbC sample is a Dirac type-II semimetal. The DFT results are in direct accordance with the ARPES measurements given in a recent report on this compound [24].

The first principle calculation of band structure shows that the NbC system respects the Time-Reversal Symmetry (TRS) since each band splits into two when SOC is included, as shown in fig.3(a,b). Moreover, the Chern number is not a good quantity to characterize the topology as it turns out to be zero for TRS systems, while the $Z_2$-invariants are found to be more appropriate for TRS systems [37]. We follow the Soulyanov-Vanderbilt [37] method of Wannier Charge Centers (WCC) that are calculated from MLWFs. These WCC are basically evolved in 6 Brillouin zone planes, which are resembled through $K_1$, $K_2$, $K_3 = 0$ and $K_1$, $K_2$, $K_3 = 0.5$. Based on these planes, the presence of Wilson's loop is determined in order to figure out the topology in the system. It can be understood through the literature [37] that an even number of the crossing of WCC implies a topologically trivial state($z2=0$), whereas an odd number of crossings indicate the presence of a topologically non-trivial state($z2=1$). The $Z_2$ topological invariant for these six planes in NbC (fig.4) are found to be

(i)      k1=0.0, k2-k3 plane: Z2= 1

(ii)     k1=0.5, k2-k3 plane: Z2= 1

(iii)    k2=0.0, k1-k3 plane:  Z2= 0

(iv)    k2=0.5, k1-k3 plane: Z2= 1

(v)     k3=0.0, k1-k2 plane: Z2= 0

(vi)    k3=0.5, k1-k2 plane: Z2= 1.

The topological Z2 index is represented as $(v_0; v_1 v_2 v_3)$. The last three Z2 numbers are the weak index, whereas the first one is the strong index. The strong index has some redundancy [37] as we can see that for the first pair of the plane (i) and (ii), the $v_0 = 0$ indicating topologically trivial, but for the other two pairs of planes, it is $v_0 = 1$, which indicates a topologically non-trivial state. Here, the weak index has no redundancy, and it is $(v_0; 111)$. Thus, weak index indicates that NbC has non-trivial topological states, and it can be surely sorted that at least, weak topology is present in NbC. Moreover, it is essential to highlight here that the Wannier charge calculation does not converge due to the presence of nodal points in the NbC system when implemented in WANNIERTOOLS as confirmed through the output file of the tool. The presence of nodal points also confirms that the NbC has a non-trivial topology. Thus, the DFT calculations, along with the Z2 invariants and Nodal points, the Non-trivial band topology in NbC and tunable superconductivity with the elevated temperature have been confirmed.

**Conclusion:**

Summarily, we have synthesized a polycrystalline sample of NbC in a single step at a higher temperature than previously reported. A non-monotonic relationship is observed in $T_c$ with lattice parameters and synthesis temperature. FC and ZFC plots confirm bulk superconductivity at 8.9K. Nature of Superconductivity is found to be strong type-II through M-H plots and phase diagrams. Simultaneously, the DFT calculations show that bands at high symmetry points do split when SOC is included which is also confirmed by calculating Z2 invariant for the same. It suggests that NbC has a topologically non-trivial band structure. This result, along with observed superconductivity at around 8.9K, suggests that NbC is a potential candidate to explore topological superconductivity.

**Acknowledgment:**

The authors would like to thank Director NPL for his keen interest and encouragement. The authors are Mr. Krishna Kandpal for vacuum encapsulation of the sample. M.M. Sharma and N.K. Karn would like to thank CSIR for the research fellowship, and Prince Sharma would like to thank UGC for his fellowship support. N.K. Karn, M.M. Sharma, and Prince Sharma are also thankful to AcSIR for Ph.D. registration.

**Table: 1**

Variation in $T_c$ and possible phase formation due to change in lattice constant by changing the synthesis temperature:

| Synthesis Temperature | Lattice parameter (Å) | $T_c$ (K) |
|---|---|---|
| $1150^0$C | 4.468(2) | 9.09 |
| $1200^0$C | 4.469(3) | 11.0 |
| $1250^0$C | 4.470(2) | 11.5 |
| $1350^0$C | 4.474(4) | 8.9 |

**Figure Captions:**

**Fig. 1:** Rietveld refined PXRD pattern of $1350^0$C synthesized NbC polycrystalline sample, and the inset shows the VESTA drawn unit cell of the same.

**Fig. 2(a):** FC & ZFC measurements of $1350^0$C synthesized NbC sample at 8 Oe, inset shows variation in $T_c$ with lattice parameters at different processing temperatures.

**Fig. 2(b):** Isothermal M-H plots of $1350^0$C synthesized NbC samples at 2K, 4K, 6K, 8K, and 10K, inset shows the same at 2K.

**Fig. 2(c):** Isothermal M-H plots of $1350^0$C synthesized polycrystalline NbC sample at 2K, 4K, and 6K in low magnetic field region, inset shows the exercise done to calculate the lower critical field $H_{c1}$.

**Fig. 2(d):** Phase diagram showing the superconducting, intermediate and normal state of synthesized NbC polycrystalline sample

**Fig. 3(a):** Calculated band structure and density of states (DOS) without considering the Spin-Orbit Coupling (SOC) for the synthesized NbC sample.

**Fig. 3(b):** Calculated band structure and Density of States with spin-orbit coupling (SOC) protocols of synthesized NbC sample, inset shows zoom view of band structure around Dirac point.

**Fig. 3(c):** Band structure with and without Spin-orbit coupling (SOC) around Fermi level at Γ a point.

**Fig. 4:** The evolution of wannier charge centers in the six planes of in Brillouin Zone. The Z2 invariants are (a) k1=0.0, k2-k3 plane: Z2= 1 (b) k1=0.5, k2-k3 plane: Z2= 1 (c) k2=0.0, k1-k3 plane:  Z2= 0 (d) k2=0.5, k1-k3 plane: Z2 = 1 (e) k3=0.0, k1-k2 plane: Z2 = 0 (f) k3=0.5, k1-k2 plane: Z2 = 1.

**References:**

[1]  M. Z. Hasan and C. L. Kane, Rev. Mod. Phys. **82**, 3045 (2010).

[2]  X. L. Qi and S. C. Zhang, Rev. Mod. Phys. **83**, 1057 (2011).

[3]  R. Sultana, P. Neha, R. Goyal, S. Patnaik, and V. P. S. Awana, J. Magn. Magn. Mater. **428**, 213 (2017).

[4]  D. Sharma, Y. Kumar, P. Kumar, V. Nagpal, S. Patnaik, and V. P. S. Awana, Solid State Commun. **323**, 114097 (2021).

[5]  J. E. Moore, Nature **464**, 194 (2010).

[6]  P. Sharma, M. Kumar, and V. P. S. Awana, J. Mater. Sci. Mater. Electron. **31**, 7959 (2020).

[7]  P. Sharma, M. M. Sharma, M. Kumar, and V. P. S. Awana, Solid State Commun. **319**, 114005 (2020).

[8]  M. Sato and Y. Ando, Reports Prog. Phys. **80**, 076501 (2017).

[9]  M. Leijnse and K. Flensberg, Semicond. Sci. Technol. **27**, 124003 (2012).

[10]  C. W. J. Beenakker, Annu. Rev. Condens. Matter Phys. **4**, 113 (2013).

[11]  S. Das Sarma, M. Freedman, and C. Nayak, Npj Quantum Inf. **1**, 15001 (2015).

[12]  Y. S. Hor, A. J. Williams, J. G. Checkelsky, P. Roushan, J. Seo, Q. Xu, H. W. Zandbergen, A. Yazdani, N. P. Ong, and R. J. Cava, Phys. Rev. Lett. **104**, 057001 (2010).

[13]  Shruti, V. K. Maurya, P. Neha, P. Srivastava, and S. Patnaik, Phys. Rev. B - Condens. Matter Mater. Phys. **92**, 020506 (2015).

[14]  M. M. Sharma, P. Rani, L. Sang, X. L. Wang, and V. P. S. Awana, J. Supercond. Nov. Magn. **33**, 565 (2020).

[15]  M. M. Sharma, L. Sang, P. Rani, X. L. Wang, and V. P. S. Awana, J. Supercond. Nov. Magn. **33**, 1243 (2020).

[16]  H. Chi, W. Liu, K. Sun, X. Su, G. Wang, P. Lošt'Ák, V. Kucek, Č. Drašar, and C. Uher, Phys. Rev. B - Condens. Matter Mater. Phys. **88**, (2013).

[17]  J. L. Zhang, S. J. Zhang, H. M. Weng, W. Zhang, L. X. Yang, Q. Q. Liu, S. M. Feng, X. C. Wang, R. C. Yu, L. Z. Cao, L. Wang, W. G. Yang, H. Z. Liu, W. Y. Zhao, S. C. Zhang, X. Dai, Z. Fang, and C. Q. Jin, Proc. Natl. Acad. Sci. U. S. A. **108**, 24 (2011).

[18]  P. P. Kong, J. L. Zhang, S. J. Zhang, J. Zhu, Q. Q. Liu, R. C. Yu, Z. Fang, C. Q. Jin, W.

G. Yang, X. H. Yu, J. L. Zhu, and Y. S. Zhao, J. Phys. Condens. Matter **25**, 362204 (2013).

[19] Y. Qi, P. G. Naumov, M. N. Ali, C. R. Rajamathi, W. Schnelle, O. Barkalov, M. Hanfland, S. C. Wu, C. Shekhar, Y. Sun, V. Süß, M. Schmidt, U. Schwarz, E. Pippel, P. Werner, R. Hillebrand, T. Förster, E. Kampert, S. Parkin, R. J. Cava, C. Felser, B. Yan, and S. A. Medvedev, Nat. Commun. **7**, 11038 (2016).

[20] H. J. Noh, J. Jeong, E. J. Cho, K. Kim, B. I. Min, and B. G. Park, Phys. Rev. Lett. **119**, 016401 (2017).

[21] N. P. Armitage, E. J. Mele, and A. Vishwanath, Rev. Mod. Phys. **90**, 015001 (2018).

[22] Z. Cui, Y. Qian, W. Zhang, H. Weng, and Z. Fang, Chinese Phys. Lett. **37**, 087103 (2020).

[23] T. Shang, J. Z. Zhao, D. J. Gawryluk, M. Shi, M. Medarde, E. Pomjakushina, and T. Shiroka, Phys. Rev. B **101**, 214518 (2020).

[24] D. Yan, D. Geng, Q. Gao, Z. Cui, C. Yi, Y. Feng, C. Song, H. Luo, M. Yang, M. Arita, S. Kumar, E. F. Schwier, K. Shimada, L. Zhao, K. Wu, H. Weng, L. Chen, X. J. Zhou, Z. Wang, Y. Shi, and B. Feng, Phys. Rev. B **102**, 205117 (2020).

[25] R. H. Willens, E. Buehler, and B. T. Matthias, Phys. Rev. **159**, 327 (1967).

[26] R. Zhan and X. Luo, J. Appl. Phys. **125**, (2019).

[27] K. Bhattacharjee, S. P. Pati, and A. Maity, Phys. Chem. Chem. Phys. **18**, 15218 (2016).

[28] R. Jha and V. P. S. Awana, J. Supercond. Nov. Magn. **25**, 1421 (2012).

[29] D. Yan, M. Yang, C. Wang, P. Song, C. Yi, and Y.-G. Shi, Supercond. Sci. Technol. (2021).

[30] E. K. Storms and N. H. Krikorian, J. Phys. Chem. **63**, 1747 (1959).

[31] C. P. Kempter, E. K. Storms, and R. J. Fries, J. Chem. Phys. **33**, 1873 (1960).

[32] C. S. Yadav and P. L. Paulose, New J. Phys. **11**, 103046 (2009).

[33] P. Giannozzi, O. Andreussi, T. Brumme, O. Bunau, M. Buongiorno Nardelli, M. Calandra, R. Car, C. Cavazzoni, D. Ceresoli, M. Cococcioni, N. Colonna, I. Carnimeo, A. Dal Corso, S. De Gironcoli, P. Delugas, R. A. Distasio, A. Ferretti, A. Floris, G. Fratesi, G. Fugallo, R. Gebauer, U. Gerstmann, F. Giustino, T. Gorni, J. Jia, M. Kawamura, H. Y. Ko, A. Kokalj, E. Kücükbenli, M. Lazzeri, M. Marsili, N. Marzari, F. Mauri, N. L. Nguyen, H. V. Nguyen, A. Otero-De-La-Roza, L. Paulatto, S. Poncé, D. Rocca, R. Sabatini, B. Santra, M. Schlipf, A. P. Seitsonen, A. Smogunov, I. Timrov, T. Thonhauser,

P. Umari, N. Vast, X. Wu, and S. Baroni, J. Phys. Condens. Matter **29**, 465901 (2017).

[34]  Y. Hinuma, G. Pizzi, Y. Kumagai, F. Oba, and I. Tanaka, Comput. Mater. Sci. **128**, 140 (2017).

[35]  AA Mostofi, JR Yates, G Pizzi, YS Lee, I Souza, D Vanderbilt, N Marzari, Comput. Phys. Commum. **185**, 2309 (2014).

[36]  QuanSheng Wu, ShengNan Zhang, Hai-Feng Song, Matthias Troyer, Alexey A. Soluyanov, Comput. Phy. Commum. **224**, 405 (2018).

[37]  A. A. Soluyanov and D. Vanderbilt, Phys. Rev. B **83**, 035108 (2011).

Fig. 1

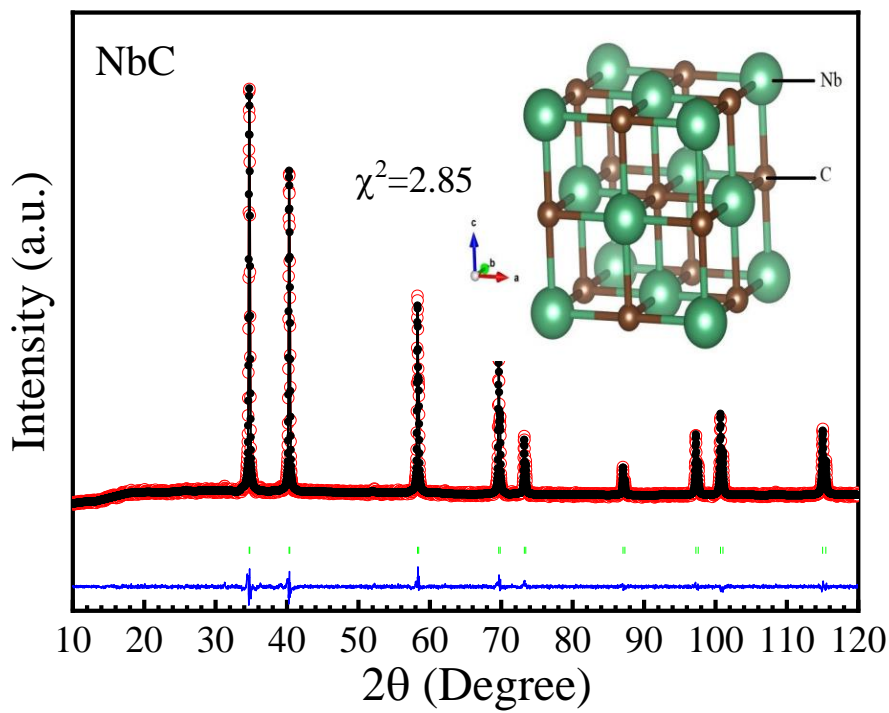

Fig. 2(a)

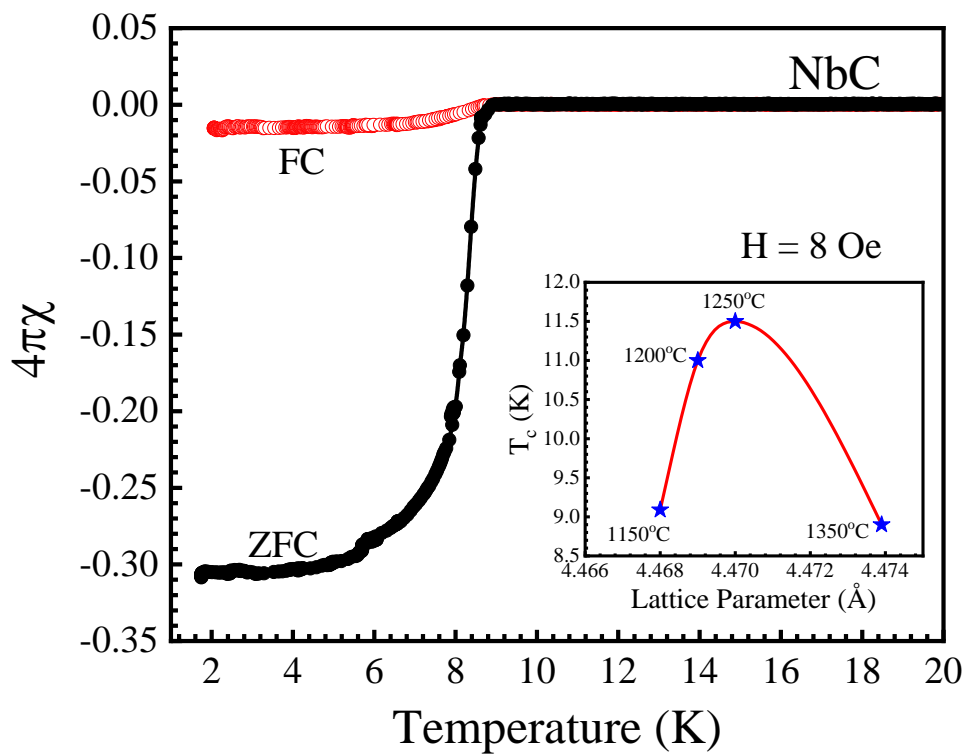

Fig. 2(b)

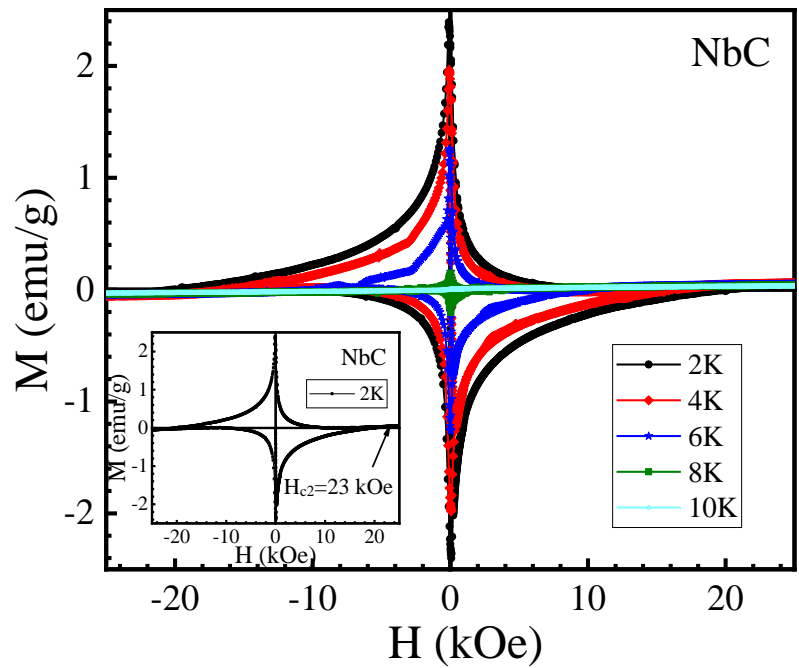

Fig. 2(c)

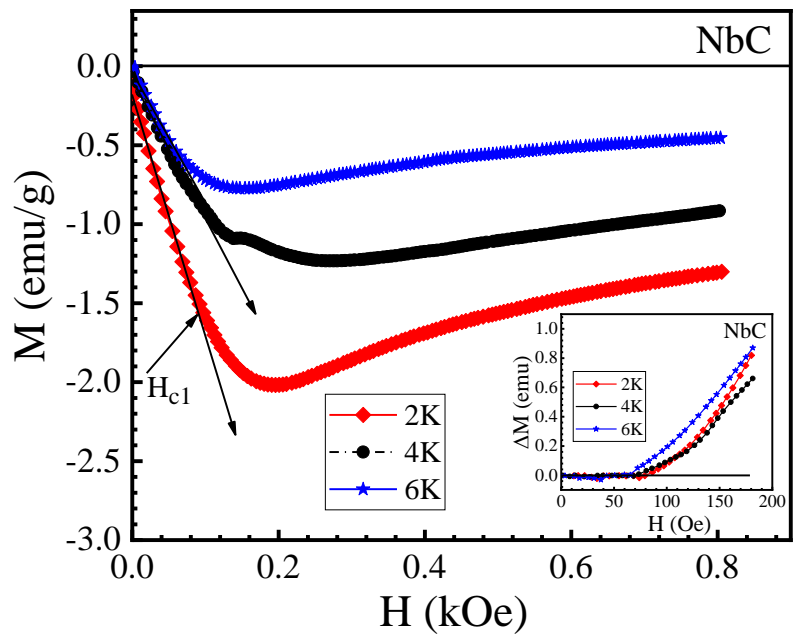

Fig. 2(d)

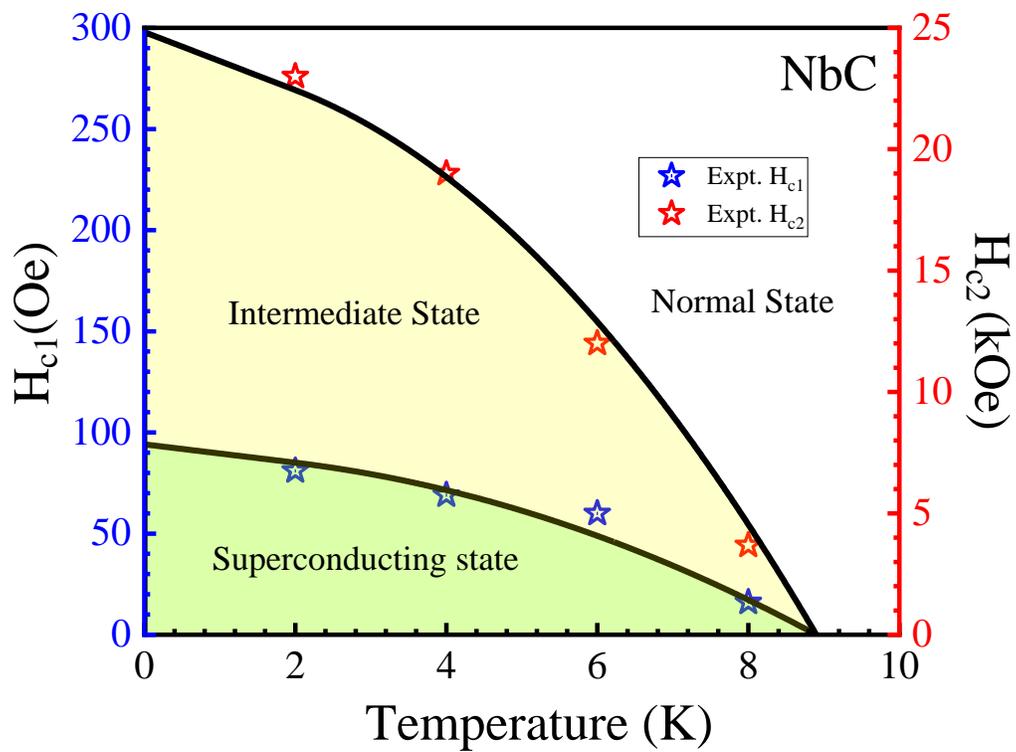

Fig. 3(a)

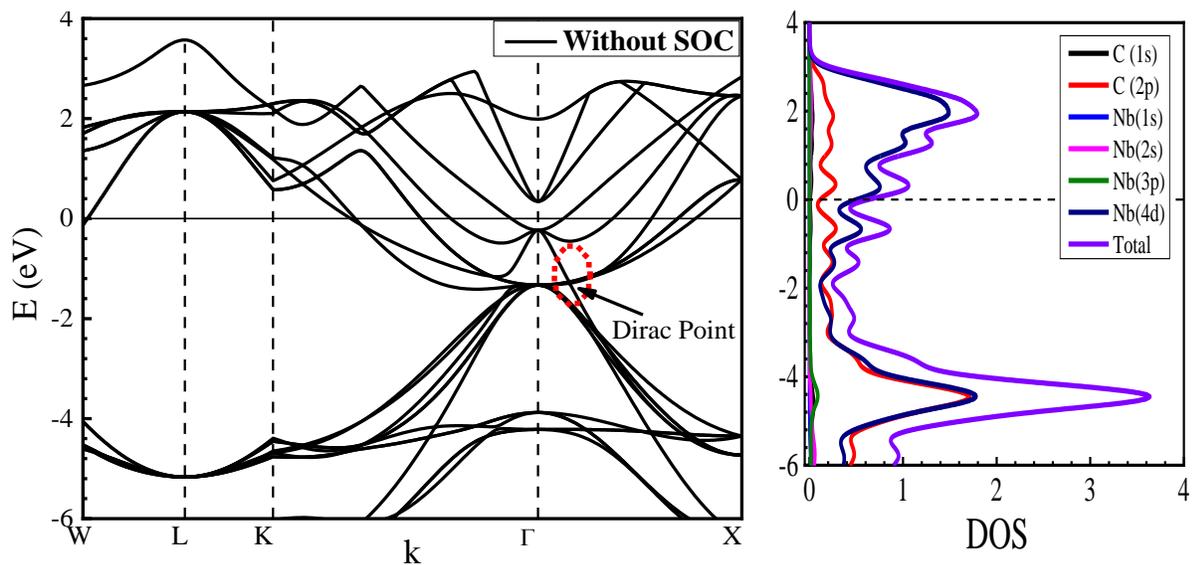

Fig. 3(b)

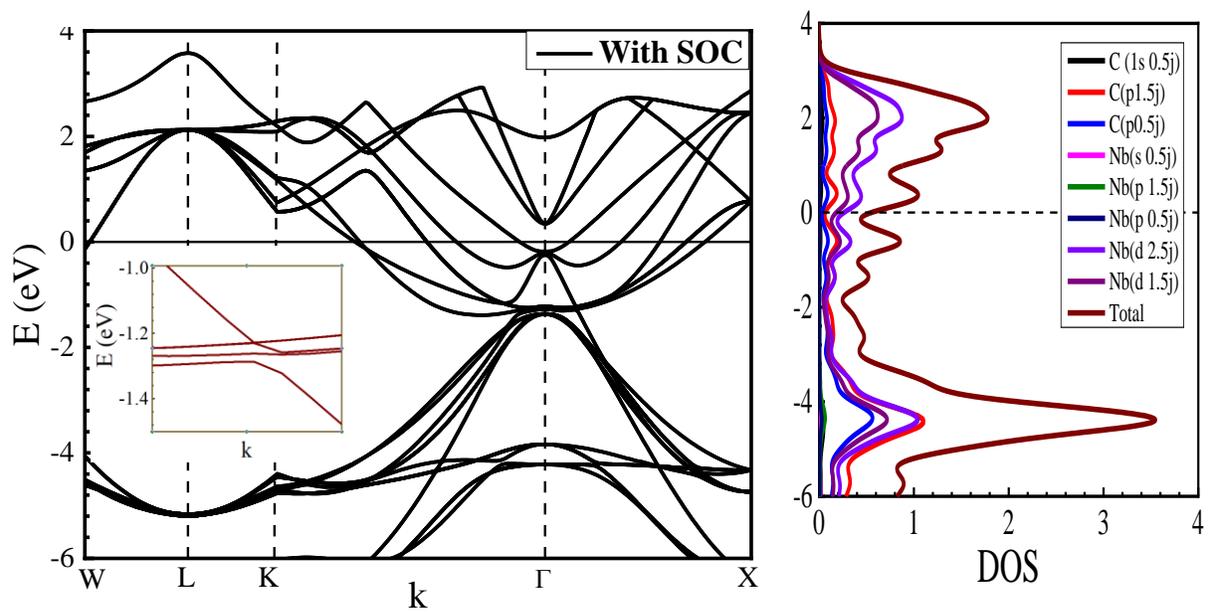

Fig. 3(c)

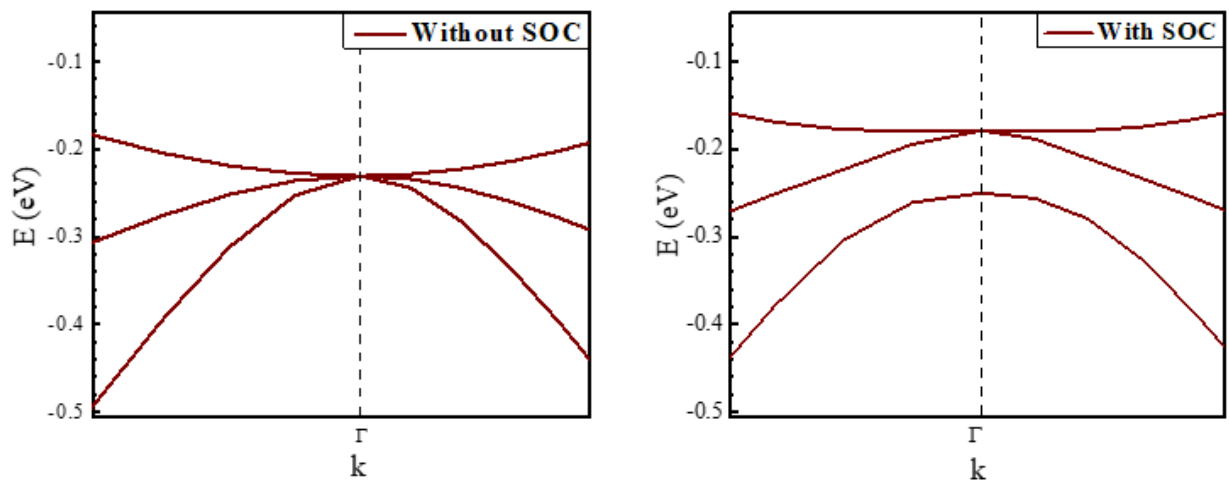

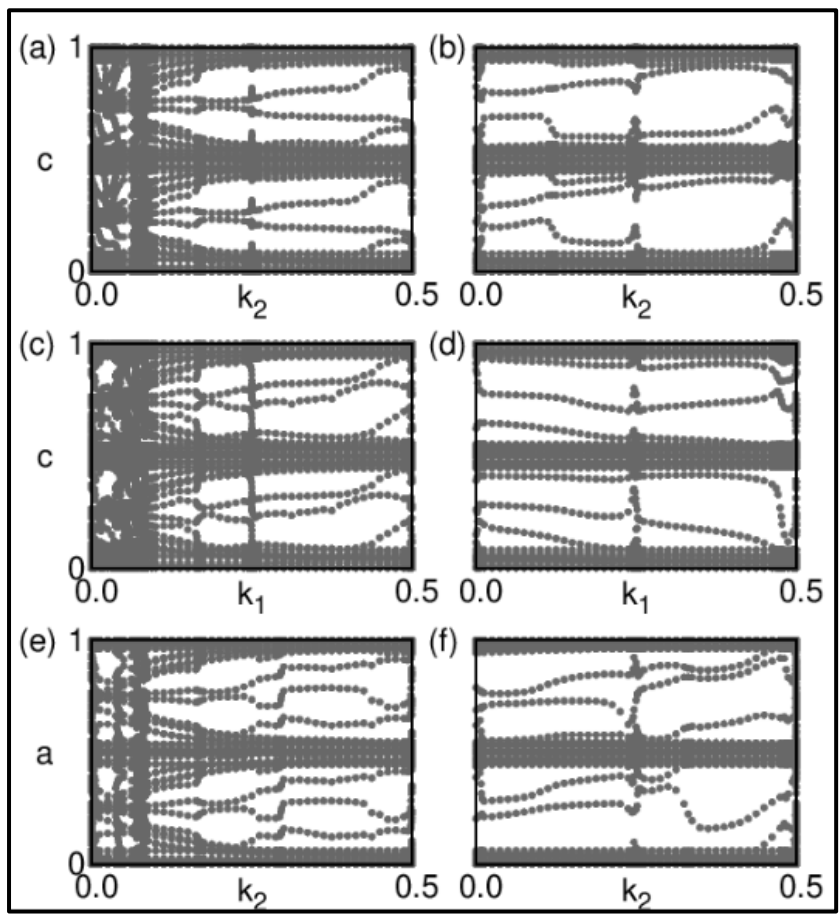